\documentclass[%
 aip,
 jmp,%
 amsmath,amssymb,
 reprint,%
 superscriptaddress,
 longbibliography
]{revtex4-1}

\usepackage{array}    
\usepackage{bm}       
\usepackage{dcolumn}  
\usepackage{float}    
\usepackage{graphicx} 
\usepackage{hhline}   
\usepackage{mhchem}   
\usepackage{multirow}
\usepackage{pbox}     
\usepackage{xr}       

\usepackage{newfloat}

\DeclareFloatingEnvironment[fileext=frm,placement={!ht},name=ALGORITHM]{algorithm}

\usepackage{titlesec}
\titlespacing*{\subsection}
{0pt}{2.5ex}{0.5ex}
\titlespacing*{\section}
{0pt}{3.5ex}{1.0ex}

\newcolumntype{C}{>{$}c<{$}}
\begin{document}

\title{The Poisson--Boltzmann model for implicit solvation of electrolyte solutions: Quantum chemical implementation and assessment via Sechenov coefficients}

\author{Christopher J. Stein}
\email[Corresponding author: ]{cjstein@lbl.gov}
\affiliation{ 
Department of Chemistry, University of California, Berkeley, California 94720 USA 
}
\affiliation{
Chemical Sciences Division, Lawrence Berkeley National Laboratory, Berkeley, California 94720 USA
}
\author{John M. Herbert}
\email[Corresponding author: ]{herbert@chemistry.ohio-state.edu}
\affiliation{ 
Department of Chemistry and Biochemistry, The Ohio State University, Columbus, Ohio 43210 USA
}
\author{Martin Head-Gordon}
\email[Corresponding author: ]{mhead-gordon@lbl.gov}
\affiliation{ 
Department of Chemistry, University of California, Berkeley, California 94720 USA 
}
\affiliation{
Chemical Sciences Division, Lawrence Berkeley National Laboratory, Berkeley, California 94720 USA
}

\begin{abstract}
We present the theory and implementation of a Poisson--Boltzmann implicit solvation model for electrolyte solutions.
This model can be combined with arbitrary electronic structure methods that provide an accurate charge density of the solute.
A hierarchy of approximations for this model includes a linear approximation for weak electrostatic potentials, finite size of the mobile electrolyte ions and a Stern-layer correction.
Recasting the Poisson--Boltzmann equations into Euler--Lagrange equations then significantly simplifies the derivation of the free energy of solvation for these approximate models.
The parameters of the model are then either fit directly to experimental observables --- e.g. the finite ion size --- or optimized for agreement with experimental results.
Experimental data for this optimization is available in the form of Sechenov coefficients that describe the linear dependence of the salting-out effect of solutes with respect to the electrolyte concentration.
In the final part we rationalize the qualitative disagreement of the finite ion size modification to the Poisson--Boltzmann model with experimental observations by taking into account the electrolyte concentration dependence of the Stern layer.
A route towards a revised model that captures the experimental observations while including the finite ion size effects is then outlined.
This implementation paves the way for the study of electrochemical and electrocatalytic processes of molecules and cluster models with accurate electronic structure methods.

\end{abstract}

\maketitle

\setlength{\parindent}{0cm}
\setlength{\parskip}{0.6em plus0.2em minus0.1em}

\section{Motivation} 

In contrast to the \textit{ab initio} calculation of chemical processes in the gas phase, the investigation of such processes in solution adds an additional layer of tremendous complexity.
The electrostatic interaction between solvent and solute directly alters the enthalpic landscape whereas osmotic pressure and reorientation of the solvent molecules based on the electrostatic interactions has a profound influence on the free energy.
Although these effects can in principle be calculated with molecular dynamics (MD) simulations, the amount of sampling required to converge the free energy combined with the large number of structures involved in chemical processes for which such a sampling would have to be performed, renders this approach unfeasible at present.\\
Hence, implicit solvation models, rather than explicit ones, are usually the method of choice in computational quantum chemistry.
Famous examples of the former are the polarizable continuum model (PCM)\cite{miertus1981,cammi1995,truong1995,cossi1996} or the conductor like screening model (COSMO),\cite{klamt1993} that exist in various flavours and specializations, are available in most multi-purpose electronic structure packages and have proven suitable in countless applications.\cite{cramer1999,tomasi2005}\\
For the description of electrochemical processes, however, the effect of the electrolyte ions needs to be taken into account.
In the realm of electrocatalysis, the phenomenological description of the solvent/electrolyte interaction with charged surfaces has a long history.
Helmholtz was the first to describe the formation of a layer of counterions in a certain distance to a charged surface --- the Helmholtz layer --- which resembles a capacitor and stores electrostatic energy.\cite{helmholtz1853}
A little later, Gouy\cite{gouy1910} and independently Chapman\cite{chapman1913} introduced the concept of a diffuse electrochemical double-layer based on the notion of an exponential decay of the electric potential away from the charged surface.
A combination of both approaches, a rigid inner layer close to the charged surface, called the Stern layer, and the Gouy--Chapman type diffuse layer, was introduced in 1924 by Stern.\cite{stern1924}
The Poisson--Boltzmann model\cite{weisbuch1981,sharp1990,sharp1995,chen1997,weetman1997,borukhov1997,harries1998,borukhov00,fogolari2002,boschitsch2011,womack2018} for implicit electrolyte solvation is a computational realization of the Gouy--Chapman model with the possibility to include a Stern-layer correction.\cite{ringe2016}
In this model, the electrostatic potential of the system is calculated from the Poisson equation appended by an additional term describing the charge density of the mobile electrolyte ions.
The charge density of the mobile ions is then modeled by a Boltzmann distribution mediated by the electrostatic potential, hence the term Poisson--Boltzmann (PB) model.
Since the electrolyte charge density depends on the total electrostatic potential that is in turn determined by the PB equations, a self-consistent solution, often on a real-space grid, is required.
The PB model is especially popular in MD simulations and biochemical applications\cite{baker2001,chu2007,yap2010,wang2014,felberg2017} but has also found increasing interest in the modeling of chemical processes on surfaces, mostly in software packages for extended solid systems employing periodic boundary conditions.\cite{jinnouchi2008,mathew2014,ringe2016}

In this article, we present a three-dimensional real-space grid implementation in the general purpose electronic structure package Q-Chem.\cite{shao2015}
This implementation paves the way to investigate chemical processes and reactions in electrolyte solutions for isolated molecules or small cluster models.
Especially for the study of electrocatalytic reactions, such a solvent model is imperative in order to calculate qualitatively meaningful free energy profiles.

We will first describe the details of the theory and the hierarchy of approximations, before we analyze the choice of parameters of the model and optimize those with respect to experimental data.
In the final section, we discuss the effect of an electrolyte concentration dependent Stern-layer thickness on the PB model and suggest future improvements on the model informed by either experiment or MD simulations.

\section{Theory}

The theory of the Poisson--Boltzmann implicit solvation model has been introduced and reviewed in numerous publications (see e.g. Refs.~\citenum{sharp1990,borukhov00,boschitsch2011,fisicaro2016,ringe2016,womack2018}).
However, the specific parameterization of our implementation and the hierarchy of approximations we introduce here requires some theoretical background that we provide in the following sections.

\subsection{Derivation of the electrostatic energy}
The total electrostatic potential $\phi^\mathrm{tot}(\mathbf{r})$ of a molecular system embedded in an electrolyte solution can be obtained as the solution to Poisson's equation appended by a term describing the charge density of the electrolyte ions $\rho^\mathrm{ions}(\mathbf{r})$
\begin{align}
\nabla[\epsilon(\mathbf{r}) \nabla \phi^\mathrm{tot}(\mathbf{r})] + 4 \pi [\rho^\mathrm{sol}(\mathbf{r}) + \rho^\mathrm{ions}(\mathbf{r})]&= 0 \nonumber \\ 
\nabla[\epsilon(\mathbf{r}) \nabla \phi^\mathrm{tot}(\mathbf{r})] + f(\phi^\mathrm{tot}(\mathbf{r}), \mathbf{r}) &= 0 \, 
\label{eq_pb}
\end{align} 

where $\epsilon (\mathbf{r})$ is the spatially dependent dielectric permittivity, $\rho^\mathrm{sol}(\mathbf{r})$ is the charge distribution of the solute, and the explicit expressions for $f(\phi^\mathrm{tot}(\mathbf{r}), \mathbf{r})$ define the different approximations we will discuss in the following. 
Following the implementation of the Poisson equation solver in the Q-Chem program package by one of us,\cite{coons2018} we define the spatially dependent dielectric constant as a product of error functions for each atom
\begin{align}
\epsilon(\mathbf{r}) = \epsilon_0 + (\epsilon_\mathrm{solv}- \epsilon_0)\prod_\alpha^\mathrm{atoms} s_\alpha^\mathrm{diel}(d_\alpha, \Delta; |\mathbf{r} -\mathbf{R}_\alpha|)
\label{eq:dielectric}
\end{align}
with the dielectric permittivity in vacuum $\epsilon_0$, the dielectric permittivity of the solvent $\epsilon_\mathrm{solv}$ and
\begin{align}
s_\alpha^\mathrm{diel}(d_\alpha, \Delta; |\mathbf{r} -\mathbf{R}_\alpha|) = \frac{1}{2} \left[ 1+\mathrm{erf}\left( \frac{|\mathbf{r} -\mathbf{R}_\alpha|-d_\alpha}{\Delta}\right)\right] \,
\label{errfunc}
\end{align}
where $d_\alpha$ is a given atom-specific radius (\textit{vide infra}) and $\Delta$ defines the interpolation length between the vacuum permittivity in the solvent region and the bulk solvent permittivity at larger distances.
The error function interpolates smoothly between these two values over a region of about $4 \Delta$.

If  $\rho^\mathrm{ions}(\mathbf{r})$ is assumed to be defined by a Boltzmann distribution of the electrolyte charges, Eq.~(\ref{eq_pb}) becomes the non-linear Poisson--Boltzmann (PB) equation with
\begin{align}
f_\mathrm{PB} (\phi,\mathbf{r}) = 4 \pi \rho^\mathrm{sol}(\mathbf{r}) + 4 \pi \lambda(\mathbf{r}) \sum_i^m q_i c_i^b \exp{\left(- \frac{q_i \phi^\mathrm{tot}(\mathbf{r})}{k_\mathrm{B}T} \right)} \,
\end{align}

where $m$ is the number of electrolyte ion species, $q_i$ is the charge of electrolyte $i$, $c_i^b$ is its bulk concentration, $k_\mathrm{B}$ is the Boltzmann constant and $T$ is the absolute temperature.
The ion exclusion function $\lambda(\mathbf{r})$ ensures that the electrolyte ion concentration tends to zero inside the solute cavity and is defined in a similar way to $\epsilon(\mathbf{r})$ with an additional parameter $a$ that corrects for the Stern layer by excluding an even larger region around the solute
\begin{align}
\lambda(\mathbf{r}) = \prod_\alpha^\mathrm{atoms}  \frac{1}{2} \left[ 1+\mathrm{erf}\left( \frac{|\mathbf{r} -\mathbf{R}_\alpha|-d_\alpha-a}{\Delta}\right)\right] \, .
\label{eq:lambda}
\end{align}

For a 1:1 electrolyte ($q_1 =  e$, $q_2 =  -e$, $c_1^b = c_2^b = c^b$, with the elementary charge $e$), the charge density of the electrolyte ions can be written as
\begin{align}
\rho^\mathrm{ions}_\mathrm{PB}(\mathbf{r})  = &\lambda(\mathbf{r}) e c^b \left[ \mathrm{exp}\left(- \frac{e \phi^\mathrm{tot}(\mathbf{r})}{k_\mathrm{B}T} \right)- \mathrm{exp}\left(\frac{e \phi^\mathrm{tot}(\mathbf{r})}{k_\mathrm{B}T} \right)\right] \nonumber \\
= & -2\lambda(\mathrm{r}) e c^b \mathrm{sinh}\left(\frac{e \phi^\mathrm{tot}(\mathbf{r})}{k_\mathrm{B}T} \right) \, .
\label{eq:rho_pb}
\end{align}

The function $f_\mathrm{PB} (\phi^\mathrm{tot}(\mathbf{r}),\mathbf{r}) $ then reduces to
\begin{align}
f_\mathrm{PB} (\phi^\mathrm{tot}(\mathbf{r}),\mathbf{r}) &= 4 \pi \rho^\mathrm{sol}(\mathbf{r})  -8\pi \lambda(\mathbf{r}) e c^b \mathrm{sinh}\left( \frac{e \phi^\mathrm{tot}(\mathbf{r})}{k_\mathrm{B}T}\right) \, .
\end{align}

The Poisson--Boltzmann equation can be recast into an Euler--Lagrange equation\cite{sharp1990} for a Lagrangian $L(\phi^\mathrm{tot},\mathbf{r}, \phi_x, \phi_y, \phi_z)$ (with $\phi_\alpha = \partial \phi / \partial \alpha$) that is still to be defined
\begin{align}
\frac{\partial L}{\partial \phi} - \sum_{\alpha \in \{x,y,z\}} \frac{\partial}{\partial \alpha}\left(\frac{\partial L}{\partial \phi_\alpha}\right) = 0 \, .
\end{align}
It follows from a theorem of the calculus of variations that the solution to this set of differential equations defines the minimum of the integral of $L$ over the independent variables $\mathbf{r}=x,y,z$.
Since the minimum condition defines the thermodynamic equilibrium state, this integral describes the electrostatic free energy of the system
\begin{align}
G^\mathrm{es} = \int L(\phi^\mathrm{tot},\mathbf{r}, \phi_x, \phi_y, \phi_z) \mathrm{d} \mathbf{r}\, .
\end{align}

In case of the non-linear Poisson--Boltzmann equation, integration gives the functional $L$ for a 1:1 electrolyte as
\begin{align}
L_\mathrm{PB} =  &4 \pi \rho^\mathrm{sol}(\mathbf{r})\phi^\mathrm{tot}(\mathbf{r}) - 8 \pi c^b k_\mathrm{B} T \lambda(\mathbf{r})\mathrm{cosh}\left( \frac{e \phi^\mathrm{tot}(\mathbf{r}) }{k_\mathrm{B} T}\right) \nonumber \\
&-\frac{\epsilon(\nabla \phi^\mathrm{tot}(\mathbf{r}))^2}{2} \, .
\end{align}

The second term describes the excess osmotic pressure due to the presence of the electrolyte ions.
The Euler-Lagrange equations are fulfilled for all $L' = C_1 L + C_0$, with the constants $C_1$ and $C_0$ that can be chosen to fulfill certain criteria for the units and boundary condition, respectively.
We therefore fix $C_1 = (4 \pi)^{-1}$ which is essentially a transformation to MKS units and $C_0 =0$ such that in the absence of $\rho^\mathrm{sol}(\mathbf{r})$ (and hence when $\phi^\mathrm{tot}(\mathbf{r})=0$) only the excess osmotic pressure term deviates from zero.

The electrostatic free energy can then be written as
\begin{align}
G^\mathrm{es}_\mathrm{PB} =& \int \rho^\mathrm{sol}(\mathbf{r})\phi^\mathrm{tot}(\mathbf{r})- 2 c^b k_\mathrm{B} T \lambda(\mathbf{r})  \mathrm{cosh}\left( \frac{e\phi^\mathrm{tot}(\mathbf{r})}{k_\mathrm{b} T}\right) \nonumber \\
&-\frac{\epsilon(\mathbf{r})(\nabla \phi^\mathrm{tot}(\mathbf{r}))^2}{8\pi} \mathrm{d}\mathbf{r} \nonumber \\
=& \int \rho^\mathrm{sol}(\mathbf{r})\phi^\mathrm{tot}(\mathbf{r})-\Delta \Pi_\mathrm{PB} - \frac{\mathbf{E}\cdot\mathbf{D}}{8 \pi} \mathrm{d}\mathbf{r}\, ,
\label{eq:G_es_pb}
\end{align}
where $\Delta \Pi_\mathrm{PB}$ is the excess osmotic pressure, $\mathbf{E} = \nabla \phi^\mathrm{tot}(\mathbf{r})$ is the electric field and $\mathbf{D}= \epsilon(\mathbf{r}) \mathbf{E}$ is the electric displacement field.

Integrating Eq.~(\ref{eq_pb}) by parts we obtain\cite{fisicaro2016}
\begin{align}
\int \frac{\mathbf{E}\cdot\mathbf{D}}{8 \pi} \mathrm{d}\mathbf{r} = \int \frac{(\rho^\mathrm{sol}(\mathbf{r}) + \rho^\mathrm{ions}(\mathbf{r}))\phi^\mathrm{tot}(\mathbf{r})}{2} \mathrm{d}\mathbf{r} \, ,
\end{align}
and the electrostatic free energy can then be written in the more common form\cite{boschitsch2011}
\begin{align}
&G^\mathrm{es}_\mathrm{PB} = \nonumber \\
 &\int  \left( \frac{1}{2} \rho^\mathrm{sol}(\mathbf{r})\phi^\mathrm{tot}(\mathbf{r})-\frac{1}{2} \rho^\mathrm{ions}(\mathbf{r})\phi^\mathrm{tot}(\mathbf{r}) -\Delta \Pi_\mathrm{PB} \right) \mathrm{d}\mathbf{r}\, .
\label{eq:elstat_short}
\end{align}

The electrostatic free energy expression is drastically simplified in the linear Poisson--Boltzmann (LPB) equation where the exponential term in the electrolyte ion charge density distribution is approximated by the first two terms of the Taylor expansion
\begin{align}
\rho^\mathrm{ions}_\mathrm{LPB}(\mathbf{r}) = \lambda(\mathbf{r}) \sum_i^m q_i c_i^b \left[1 - \frac{q_i \phi^\mathrm{tot}(\mathbf{r})}{k_\mathrm{B} T}\right] \, .
\end{align}

Restricting ourselves again to the case of a 1:1 electrolyte this reduces to
\begin{align}
\rho^\mathrm{ions}_\mathrm{LPB}(\mathbf{r}) = -2 \lambda(\mathbf{r}) \frac{e^2c^b}{k_\mathrm{B} T} \phi^\mathrm{tot}(\mathbf{r}) \, .
\label{eq:rho_lpb}
\end{align}

The functional $L_\mathrm{LPB}$ in the linear case can then be written as
\begin{align}
L_\mathrm{LPB} = &4 \pi \rho^\mathrm{sol}(\mathbf{r})\phi^\mathrm{tot}(\mathbf{r}) - \frac{4\pi e^2c^b}{k_\mathrm{B} T} \lambda(\mathbf{r}) (\phi^{\mathrm{tot}}(\mathbf{r}))^2\nonumber \\
&- \frac{\epsilon(\nabla \phi^\mathrm{tot}(\mathbf{r}))^2}{8 \pi} \nonumber \\
=& 4 \pi \rho^\mathrm{sol}(\mathbf{r})\phi^\mathrm{tot}(\mathbf{r}) + 4 \pi \frac{\rho^\mathrm{ions}(\mathbf{r})\phi^\mathrm{tot}(\mathbf{r})}{2} \nonumber \\
&- \frac{\epsilon(\nabla \phi^\mathrm{tot}(\mathbf{r}))^2}{8\pi} \, .
\end{align}

With the same arguments as above we can fix the constants $C_0 = 0$ and $C_1=1/4\pi$.
The electrostatic free energy for the linear Poisson--Boltzmann equation then reads
\begin{align}
G^\mathrm{es}_\mathrm{LPB} = & \int \left( \rho^\mathrm{sol}(\mathbf{r}) \phi^\mathrm{tot}(\mathbf{r})+\frac{1}{2} \rho^\mathrm{ions}(\mathbf{r})\phi^\mathrm{tot}(\mathbf{r}) -  \frac{\mathbf{E}\cdot\mathbf{D}}{8 \pi}\right) \mathrm{d} \mathbf{r} \nonumber \\
=&\int \frac{1}{2} \rho^\mathrm{sol}(\mathbf{r}) \phi^\mathrm{tot}(\mathbf{r}) \mathrm{d} \mathbf{r} \, ,
\label{eq:elstat_en_lin}
\end{align}
which is the same expression as for a pure solvent without electrolyte and only the electrostatic potential is modified by the presence of the electrolyte ions.

\subsection{Modified Poisson--Boltzmann equation}

The standard Poisson--Boltzmann equation assumes point-like ions which leads to a charge accumulation of the electrolyte that is unphysically large.
Finite ion size can however be included in the theory leading to the modified Poisson--Boltzmann (MPB) equation.\cite{borukhov1997,boschitsch2011,ringe2016,castaneda2019}
This can be achieved by a modification of the expression for the ionic concentrations
\begin{align}
c_i(\mathbf{r}) = \frac{\lambda(\mathbf{r})c_i^b\mathrm{exp}\left( -\frac{q_i \phi^\mathrm{tot}(\mathbf{r})}{k_\mathrm{B}T}\right)}{1 + \sum_{j=1}^m \frac{c_j^b}{c_j^\mathrm{max}}\left[ \lambda(\mathbf{r})\mathrm{exp}\left( -\frac{q_j \phi^\mathrm{tot}(\mathbf{r})}{k_\mathrm{B}T}\right)-1\right]} \, .
\end{align}

The highest possible concentration for electrolyte species $j$ assuming closest-packing $c_j^\mathrm{max}$ can be written as
\begin{align}
c_j^\mathrm{max} = \frac{p}{\frac{4}{3} \pi N_\mathrm{A}R_j^3}\, ,
\end{align}
with the closest-packing factor $p=0.74$, the Avogadro number $N_\mathrm{A}$, and the effective ion radius of the electrolyte ions $R_j$.
Restricting ourselves again to a 1:1 electrolyte we obtain for the charge density of the electrolyte ions
\begin{align}
\rho^\mathrm{ions}_\mathrm{MPB}(\mathbf{r}) &= -\frac{2\lambda(\mathbf{r}) e c^b \mathrm{sinh}\left( \frac{e\phi^\mathrm{tot}(\mathbf{r})}{k_\mathrm{B} T}\right)}{1 + \sum_{j=1}^m \frac{c_j^b}{c_j^\mathrm{max}}\left[ \lambda(\mathbf{r})\mathrm{exp}\left( -\frac{q_j \phi^\mathrm{tot}(\mathbf{r})}{k_\mathrm{B}T}\right)-1\right]}\nonumber \\
& = - \frac{2\lambda(\mathbf{r}) e c^b \mathrm{sinh}\left( \frac{e\phi^\mathrm{tot}(\mathbf{r})}{k_\mathrm{B} T}\right)}{1 - \frac{c^b}{c_{1+2}}+\frac{c^b}{c_{1+2}}\lambda(\mathbf{r})\mathrm{cosh}\left( \frac{e \phi^\mathrm{tot}(\mathbf{r})}{k_\mathrm{B}T}\right)} \, ,
\label{eq:rho_mpb}
\end{align}
with
\begin{align}
c_{1+2} = \frac{p}{\frac{4}{3}\pi N_\mathrm{A}\left( R_1^3+R_2^3\right)} \, .
\end{align}

Integration with respect to $\phi^\mathrm{tot}(\mathbf{r})$ gives the functional
\begin{align}
L_\mathrm{MPB} =  &4 \pi \rho^\mathrm{sol}(\mathbf{r})\phi^\mathrm{tot}(\mathbf{r}) -\frac{\epsilon(\nabla \phi^\mathrm{tot}(\mathbf{r}))^2}{2} -8 \pi k_\mathrm{B}T c_{1+2} \nonumber \\
& \times\mathrm{ln}\left[1+ \frac{c^b}{c_{1+2}}\left(\lambda(\mathbf{r}) \mathrm{cosh}\left(\frac{e\phi^\mathrm{tot}(\mathbf{r})}{k_\mathrm{B}T}\right) -1\right)\right]
\end{align}

Proceeding in the same manner as above we obtain the constants $C_1 = 1/4\pi$ and $C_0 = 0$ leading to the following expression for the electrostatic energy
\begin{align}
&G^\mathrm{es}_\mathrm{MPB} = \int \rho^\mathrm{sol}(\mathbf{r})\phi^\mathrm{tot}(\mathbf{r})-\frac{\epsilon(\mathbf{r})(\nabla \phi^\mathrm{tot}(\mathbf{r}))^2}{8\pi} \nonumber \\
&-2 k_\mathrm{B} T c_{1+2} \mathrm{ln}\left[1+ \frac{c^b}{c_{1+2}}\left(\lambda(\mathbf{r}) \mathrm{cosh}\left(\frac{e\phi^\mathrm{tot}(\mathbf{r})}{k_\mathrm{B}T}\right) -1\right)\right]\mathrm{d}\mathbf{r} \nonumber \\
&= \int \rho^\mathrm{sol}(\mathbf{r})\phi^\mathrm{tot}(\mathbf{r})-\Delta \Pi_\mathrm{MPB} - \frac{\mathbf{ED}}{8 \pi} \mathrm{d}\mathbf{r}\, ,
\label{eq:G_es_mpb}
\end{align}
or alternatively Eq.~(\ref{eq:elstat_short}) where the excess osmotic pressure is now replaced by $\Delta \Pi_\mathrm{MPB}$.

The finite ion size also alters the expression for the electrolyte charge density for the linearized version of the modified Poisson--Boltzmann equation (LMPB)
\begin{align}
&\rho^\mathrm{ions}_\mathrm{LMPB} = \nonumber \\
&-\frac{2\lambda(\mathbf{r})\frac{e^2c^b}{k_\mathrm{b}T}\phi^\mathrm{tot}(\mathbf{r})}{1-\frac{c^b}{c_{1+2}}(1-\lambda(\mathbf{r}))+ \frac{e\phi^\mathrm{tot}(\mathbf{r})}{k_\mathrm{b}T} \lambda(\mathbf{r}) c^b\left( \frac{1}{c_2^\mathrm{max}}-\frac{1}{c_1^\mathrm{max}}\right)} \nonumber \\
&\approx -\frac{2\lambda(\mathbf{r})\frac{e^2c^b}{k_\mathrm{b}T}\phi^\mathrm{tot}(\mathbf{r})}{1-\frac{c^b}{c_{1+2}}(1-\lambda(\mathbf{r}))} \, .
\label{eq:rho_ions_finite_lin}
\end{align}

Note that the last term in the denominator in the first line is usually very small and even vanishes for equal ionic radii.
The second line of Eq.~(\ref{eq:rho_ions_finite_lin}) offers a simple interpretation of the finite size effect:
The charge density is scaled by the fraction of the maximum electrolyte density ($c^b/c_{1+2}$) when the ion exclusion function $\lambda(\mathbf{r}) < 1$ meaning that far away from the solute, the charge density is unaffected by the finite size effects.
As the electrolyte charge density approaches zero close to the solute anyway, the finite ion size effect is most important close to the Stern layer.

The LMPB expression for the electrostatic energy (Eq.~(\ref{eq:elstat_en_lin})), however, is the same as for the LPB case.

\subsection{Free energies of solvation}
As the total electrostatic free energy is rarely of interest, we will focus now on expressions for the free energy of solvation in an electrolyte solution.
In general, this quantity can be expressed as
\begin{align}
\Delta G^\mathrm{solv} = \,& G^\mathrm{es}(\epsilon, c^b, \rho^\mathrm{sol}(\mathbf{r}))-G^\mathrm{es}(\epsilon=1, c^b=0, \rho^\mathrm{sol}(\mathbf{r}))\nonumber \\ &-G^\mathrm{es}(\epsilon, c^b, \rho^\mathrm{sol}(\mathbf{r})=0)\, ,
\label{eq:g_solv}
\end{align}
where the first term is the electrostatic free energy of the solute in the electrolyte solution, the second term describes the electrostatic free energy of the solute in vacuum and the last term corresponds to the pure electrolyte solution.
The absence of the charge density of the solute ($\rho^\mathrm{sol}(\mathbf{r})=0$) results in $\phi^\mathrm{tot}(\mathbf{r})=0$ and $\lambda(\mathbf{r})=0$ such that the individual terms of Eq.~(\ref{eq:g_solv}) can be written as
\begin{align}
\Delta G^\mathrm{solv} = & \int \rho^\mathrm{sol}(\mathbf{r})\phi^\mathrm{tot} (\mathbf{r})\mathrm{d}\mathbf{r} - \int \Delta \Pi_\mathrm{PB} \mathrm{d}\mathbf{r} \nonumber \\
& - \frac{1}{2} \int (\rho^\mathrm{sol}(\mathbf{r})+\rho^\mathrm{ions}(\mathbf{r}))\phi^\mathrm{tot}(\mathbf{r}) \mathrm{d}\mathbf{r} \nonumber \\
& -  \int \rho^\mathrm{sol}(\mathbf{r})\phi^\mathrm{sol}(\mathbf{r}) \mathrm{d}\mathbf{r} + \frac{1}{2} \int \rho^\mathrm{sol}(\mathbf{r})\phi^\mathrm{sol} (\mathbf{r})\mathrm{d}\mathbf{r} \nonumber \\
& + \int \Delta \Pi_\mathrm{PB}^{\lambda(\mathbf{r}) = 1, \phi^\mathrm{tot}(\mathbf{r})=0} \mathrm{d}\mathbf{r} \nonumber \\
= & \frac{1}{2}  \int \rho^\mathrm{sol}(\mathbf{r})(\phi^\mathrm{tot}(\mathbf{r})  - \phi^\mathrm{sol}(\mathbf{r})) \mathrm{d}\mathbf{r} \nonumber \\
& -  \frac{1}{2} \int \rho^\mathrm{ions}(\mathbf{r})\phi^\mathrm{tot}(\mathbf{r}) \mathrm{d}\mathbf{r} \nonumber \\
& + \int \Delta \Pi_\mathrm{PB}^{\lambda(\mathbf{r}) = 1, \phi^\mathrm{tot}(\mathbf{r})=0} \mathrm{d}\mathbf{r} - \int \Delta \Pi_\mathrm{PB} \mathrm{d}\mathbf{r}\, .
\label{eq:g_solv_detail}
\end{align}

Starting from Eq.~(\ref{eq:g_solv_detail}), we can now write down the expressions for the free energies of solvation for the different approximations.
In case of the L(M)PB, the free energy of solvation is simply
\begin{align}
\Delta G^\mathrm{solv}_\mathrm{L(M)PB} = \frac{1}{2}  \int \rho^\mathrm{sol}(\mathbf{r})(\phi^\mathrm{tot}(\mathbf{r})  - \phi^\mathrm{sol}(\mathbf{r})) \mathrm{d}\mathbf{r} \, ,
\end{align}
because $\int \Delta \Pi_\mathrm{PB}^{\lambda(\mathbf{r}) = 1, \phi^\mathrm{tot}(\mathbf{r})=0} \mathrm{d}\mathbf{r}  = 0$ and $\int \Delta \Pi_\mathrm{PB} \mathrm{d}\mathbf{r} = \frac{1}{2} \int \rho^\mathrm{ions}(\mathbf{r})\phi^\mathrm{tot}(\mathbf{r}) \mathrm{d}\mathbf{r}$.
For the PB model without size modification of the ions we have
\begin{align}
\Delta G^\mathrm{solv}_\mathrm{PB} =  & \frac{1}{2}  \int \rho^\mathrm{sol}(\mathbf{r})(\phi^\mathrm{tot}(\mathbf{r})  - \phi^\mathrm{sol}(\mathbf{r})) \mathrm{d}\mathbf{r} \nonumber \\
- & \frac{1}{2} \int \rho^\mathrm{ions}(\mathbf{r})\phi^\mathrm{tot}(\mathbf{r}) \mathrm{d}\mathbf{r} \nonumber \\
+ & 2c^b k_\mathrm{B} T \int \left(1-\lambda(\mathbf{r}) \mathrm{cosh}\left( \frac{e \phi^\mathrm{tot}(\mathbf{r})}{k_\mathrm{B} T} \right)\right)\mathrm{d}\mathbf{r}
\end{align}
and finally the expression for the MPB solvation free energy reads
\begin{align}
\Delta G^\mathrm{solv}_\mathrm{MPB}=& \frac{1}{2}  \int \rho^\mathrm{sol}(\mathbf{r})(\phi^\mathrm{tot}(\mathbf{r})  - \phi^\mathrm{sol}(\mathbf{r})) \mathrm{d}\mathbf{r} \nonumber \\
- & \frac{1}{2} \int \rho^\mathrm{ions}(\mathbf{r})\phi^\mathrm{tot}(\mathbf{r}) \mathrm{d}\mathbf{r} - c^\mathrm{max} k_\mathrm{B} T \nonumber \\
  \times &\int \mathrm{ln} \left[ 1+ \frac{2 c^b}{c^\mathrm{max}} \left( \lambda(\mathbf{r}) \mathrm{cosh} \left( \frac{e \phi^\mathrm{tot}(\mathbf{r})}{k_\mathrm{B} T}\right)-1\right) \right] \mathrm{d}\mathbf{r}\, .
\end{align}
We note here that this expression is identical to the expression for the electrostatic part of the free energy of solvation in Refs.~\citenum{borukhov00,ringe2016,ringe2017} when the packing factor for a simple cubic lattice ($p \approx 0.52$) is assumed instead of the closest-packing factor.
This corresponds to the lattice-gas model assumed in that work.

\subsection{Algorithm to solve for the total electrostatic potential}

In order to solve Eq.~(\ref{eq_pb}) we follow the algorithm described by Fisicaro \textit{et al.} in Ref.~\citenum{fisicaro2016}.
We first isolate the effect of the dielectric response into a polarization charge density $\rho^\mathrm{pol}(\mathbf{r})$
\begin{align}
\label{eq:rho_iter}
\nabla^2 \phi^\mathrm{tot}(\mathbf{r}) &= -4 \pi \frac{\rho^\mathrm{tot}(\mathbf{r})}{\epsilon(\mathbf{r})}-\underbrace{\nabla \mathrm{ln} \epsilon (\mathbf{r}) \nabla \phi^\mathrm{tot}(\mathbf{r})}_{\rho^\mathrm{iter}(\mathbf{r})}\\
\label{eq:phi_tot}
&=-4\pi \left[ \rho^\mathrm{tot}(\mathbf{r})+\underbrace{\frac{1-\epsilon(\mathbf{r})}{\epsilon(\mathbf{r})}\rho^\mathrm{tot}(\mathbf{r}) +\rho^\mathrm{iter}(\mathbf{r})}_{\rho^\mathrm{pol}(\mathbf{r})}\right]\, ,
\end{align}

with $\rho^\mathrm{tot}(\mathbf{r}) = \rho^\mathrm{sol}(\mathbf{r}) +\rho^\mathrm{ions}(\mathbf{r})$.
Starting from $\rho^\mathrm{ions}(\mathbf{r})=0$  and $\phi^\mathrm{tot}(\mathbf{r})=0$ we converge the total electrostatic potential as shown in Algorithm 1.
We observed that a much stronger damping ($\kappa = 0.2$) is necessary to converge $\rho^\mathrm{ions}(\mathbf{r})$ than is necessary for $\rho^\mathrm{iter}(\mathbf{r})$ ($\eta=0.6$).

\begin{algorithm}[b!]
\caption{Self-consistent iterative procedure to converge $\phi^\mathrm{tot}$ by converging $\rho^\mathrm{iter}$ and $\rho^\mathrm{ions}$.}
\begin{tabular}{lll}
\hline
\multicolumn{3}{l}{\textbf{repeat} $k+1$ \textbf{times} until convergence}\\
\hline
&\textbf{calculate} &$\rho^\mathrm{iter}_{k+1}(\mathbf{r})$ (right part of Eq.~\ref{eq:rho_iter})\\
&\textbf{damp} &$\rho^\mathrm{iter}_{k+1}(\mathbf{r}) = \eta \rho^\mathrm{iter}_{k+1}(\mathbf{r}) +(1-\eta)\rho^\mathrm{iter}_{k}(\mathbf{r})$\\
& & $\eta = 0.6$\\
&\textbf{update}    & $\rho^\mathrm{tot}_{k+1}(\mathbf{r}) = \rho^\mathrm{sol}(\mathbf{r}) +\rho_k^\mathrm{ions}(\mathbf{r})$\\
&\textbf{calculate} & $\phi^\mathrm{tot}_{k+1}(\mathbf{r}) $ (with multigrid solver from Eq.~\ref{eq:phi_tot})\\
&\textbf{update }   & $\rho^\mathrm{ions}_{k+1}(\mathbf{r})$ (Eqs.~\ref{eq:rho_pb}, \ref{eq:rho_lpb}, \ref{eq:rho_mpb}, or \ref{eq:rho_ions_finite_lin}) \\
&\textbf{damp} &$\rho^\mathrm{ions}_{k+1}(\mathbf{r}) = \kappa \rho^\mathrm{ions}_{k+1}(\mathbf{r}) +(1-\kappa)\rho^\mathrm{ions}_{k}(\mathbf{r})$\\
& & $\kappa = 0.2$\\
\hline
\end{tabular}
\end{algorithm}

\begin{figure*}[htb]
\includegraphics[width=\textwidth]{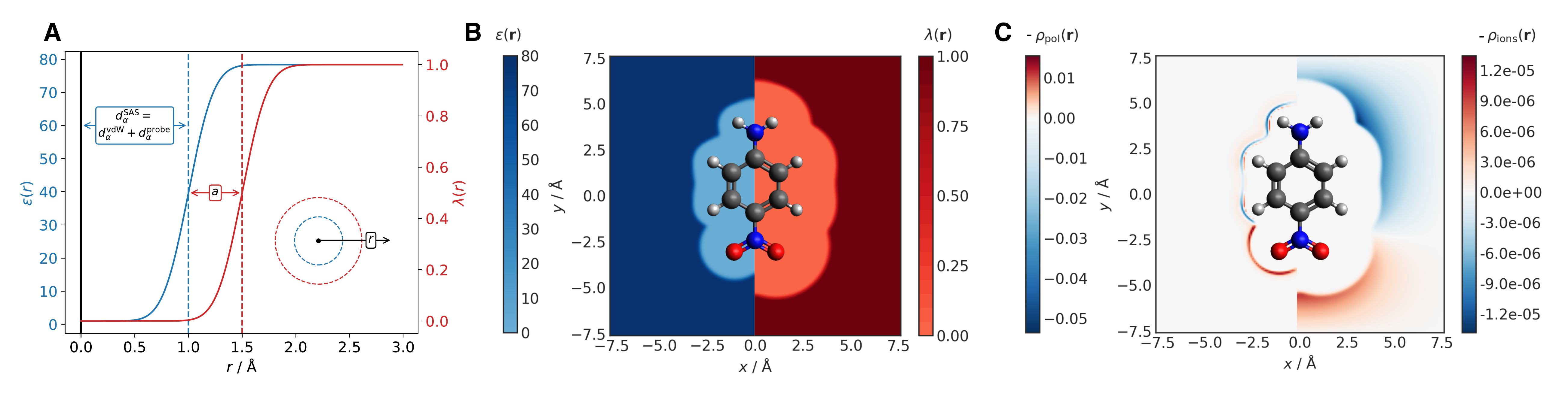}
\caption{Spatial distribution of the dielectric permittivity $\epsilon(\mathbf{r})$ Eq.~(\ref{eq:dielectric}) (blue) and the ion exclusion function $\lambda(\mathbf{r})$ Eq.~(\ref{eq:lambda}) (red). Panel A displays both functions for a single atom with $d_\alpha^\mathrm{SAS} = d_\alpha^\mathrm{vdW} +d_\alpha^\mathrm{probe} =1.0 \, \mathrm{\AA}$, a Stern layer thickness of $a=0.5\, \mathrm{\AA}$ and an interpolation length of $\Delta=0.265\, \mathrm{\AA}$. Panel B  displays $\epsilon(\mathbf{r})$ and $\lambda(\mathbf{r})$ for a cut through the molecular plane of 4-nitroaniline employing $d_\alpha = 1.2 r_\mathrm{vdW}$ and a Stern layer thickness of $a=1.0\, \mathrm{\AA}$. The polarization charge density $\rho_\mathrm{pol}(\mathbf{r})$ and the electrolyte ion charge density $\rho_\mathrm{ion}(\mathbf{r})$ are shown on the left and right hand side of panel C, respectively.}
\label{fig:aniline}
\end{figure*}

We will now discuss some of the properties of our parameterization on the example of 4-nitroaniline in 15~\AA \,cubic box (see Figure~\ref{fig:aniline}).
Panel A shows the error function for the dielectric permittivity $\epsilon(\mathbf{r})$ (blue line, left y-axis) and the ion exclusion function $\lambda(\mathbf{r})$ (red line, right y-axis) for a hypothetical atom with $d_\alpha = 1.0$ \AA \, and a Stern-layer thickness $a=0.5$ \AA.
In panel B these two functions are displayed for a cut through the main $\sigma$ symmetry plane of 4-nitroanisole.
Clearly, the solvent (blue) --- characterized solely by its dielectric permittivity --- is closer to the the solute molecule than the electrolyte ions (red) because of the Stern-layer.
The converged polarization charge density $\rho^\mathrm{pol}(\mathbf{r})$ and the ion charge density $\rho^\mathrm{ions}(\mathbf{r})$ are displayed in panel C.
As can be seen from Eq.~\ref{eq:phi_tot}, the polarization charge density resides in the domain where $\epsilon(\mathbf{r})$ varies, meaning in the onset of the error function.
For the electrolyte ion charge density, a clear separation of the positively and negatively charged ions is observed, resulting in an accumulation of positive charge close to the negatively polarized part of the solute molecule and \textit{vice versa}.

\section{Computational Details}
All calculations were carried out in a development version of Q-Chem 5.2.\cite{shao2015}
Unless otherwise noted we employed the $\omega$B97X-V density functional\cite{mardirossian2014} with a def2-TZVPP basis set\cite{weigend2005} and standard integration grids.
Our implementation builds up on a previous multigrid solver for the Poisson equation.\cite{coons2016,coons2018}
Unlike the Poisson solver that is reported in Ref.~\citenum{coons2018}, we do not apply Gaussian blurring to the nuclear contribution to the electrostatic potential.
In Section~\ref{sect:sechenov} we place the molecule in a cubic box with an edge length of 20~\AA.
With 97 grid points in each dimension the grid spacing amounts to 0.206 \AA.
After some initial tests the convergence threshold for the conjugate gradient step on each multigrid level was set to $10^{-5}$ a.u., the convergence of the polarization charge density $\rho^\mathrm{pol}(\mathbf{r})$ was set to $10^{-3}$ a.u., and the convergence for the electrolyte ion charge density $\rho_\mathrm{ion}(\mathbf{r})$ was set to $10^{-4}$ a.u. for each grid point.

The initial implementation described here is not yet as efficiently parallelized as other highly optimized multigrid solver implementations described in Refs.~\citenum{boschitsch2011,womack2018,castaneda2019}. 
Solving for the electrostatic potential hence takes considerably longer than the electronic structure calculation.
In case of caffeine, the Poisson--Boltzmann calculation takes  roughly five times longer than the electronic structure calculation.
The benefit of the real space grid implementation is certainly the independence of the calculations for individual grid points that enables excellent scaling over several threads.
In future work, we will hence focus on a more efficient MPI parallelization.
The uniform Cartesian grid employed here might also not be an optimal choice since the electrostatic potential, as well as the charge density distributions vary more close to the solute cavity then they do at the further away from the solute.
Adaptive grids\cite{boschitsch2011} or more generally adaptive mesh refinement methods,\cite{berger1984,berger1989} might be a more economical choice and can be combined with multigrid approaches\cite{brandt1977,briggs2000,brown2005} and help to focus to concentrate the computational resources to the areas where they are needed most.
For a moderate number of grid points, typical schemes for convergence acceleration of self-consistent approaches like the direct inversion of the iterative subspace (DIIS) method,\cite{pulay1980,pulay1982} can further be applied to efficiently reduce the number of iterations.
This approach can be understood as an advancement over the damping with the parameters $\eta$ and $\kappa$ in our current algorithm.

\section{Results}
Having introduced the hierarchy of PB models with their respective approximations, we will now analyze the numerical accuracy, find an optimal choice for the parameters of the model and discuss the implications upon inclusion of finite ion size.

\subsection{The Debye--H\"uckel and Kirkwood--Onsager models}

\begin{table*}[htb]
  \begin{center}
    \caption{Electrostatic free energies for the Debye--H\"uckel ion model with an ion radius $a=2.0$ \AA. The values in the last three columns were calculated from the linear Poisson--Boltzmann equation for three different interpolation lengths $l$.}
    \label{tab:bohr}
    \begin{tabular}{lllllll} 
    \hline
      $\epsilon_r$ & $\lambda$ / \AA & $c^b$ / mol l$^{-1}$ & $\Delta \Delta G_\mathrm{ion}^\mathrm{exact}$ / kcal mol$^{-1}$ & \multicolumn{3}{c}{$\Delta \Delta G_\mathrm{ion}^\mathrm{LPBE}$  / kcal mol$^{-1}$}\\
       \multicolumn{4}{c}{} & $l = 0.50$ \AA & $l = 0.35$ \AA &$l = 0.25$ \AA  \\
      \hline
 4  & 25 & $7.54 \cdot 10^{-4}$  & -1.537 &-1.214 &-1.271  &-1.236  \\
 4  & 5  & 0.0189                        & -5.935 & -5.750 &-5.838  &-5.869  \\
 4  & 3  & 0.0524                        & -8.303 & -7.922&-8.000  &-8.302  \\
 20 & 25 & $3.77 \cdot 10^{-3}$  & -0.307& -0.216& -0.323 &-0.259  \\
 20 &   5 & 0.0943                       & -1.186& -1.141&-1.208  &-1.230  \\
 20 &   3 & 0.261                         & -1.659& -1.548& -1.651 &-1.691  \\
 80 & 25 & 0.0151                       & -0.077& -0.108& -0.063  &-0.049  \\
 80 &   5 & 0.377                         & -0.296&-0.424 & -0.379 &-0.300  \\
 80 &   3 & 1.05                           & -0.415&-0.491 & -0.474 &-0.518  \\
    \hline
    \multicolumn{3}{l}{\textbf{MAE}} & &0.152 & 0.096 & 0.070   \\
    \multicolumn{3}{l}{\textbf{MRE} [\%]} && 19.0 & 10.1 & 14.2   \\
    \hline
    \end{tabular}
  \end{center}
\end{table*}

In order to assess the numerical accuracy we can expect from our model, we first calculate the electrostatic free energies of solvation for a singly charged cation with a radius of $a=2$\AA.
If the linear Poisson--Boltzmann equation is to be employed, the result for the free energy of solvation can be compared to the analytical solution provided by the Debye--H\"uckel theory\cite{debye1923,che2008,lange2011}
\begin{align}
\Delta G(a) = \frac{1}{2a}\left( \frac{1}{\epsilon_r} - \frac{1}{\epsilon_0}\right) - \frac{\kappa}{2 \epsilon_r(1+\kappa a)} \, ,
\end{align}
where $a$ is the ion radius, $\epsilon_r$ is the dielectric permittivity of the solvent and $\kappa$ is the inverse Debye screening length given by
\begin{align}
\kappa = \sqrt{\frac{4\pi \sum_i^N z_i^2 e^2 c^b_i}{\epsilon_r k_\mathrm{B} T}} \, ,
\end{align}
where $z_i$ is the integer charge of the electrolyte species $i$, $c^b_i$ is its bulk concentration, $e$ is the elementary charge, $k_\mathrm{B}$ is the Boltzmann constant and $T$ is the absolute temperature.
As in Ref.~\citenum{lange2011}, we calculate the effect of the mobile ions on the free energy of solvation 
\begin{align}
\Delta \Delta G_\mathrm{ion} = \Delta G^\mathrm{solv}(c^b) - \Delta G^\mathrm{solv}(c^b = 0)
\label{eq:delta_g_ion}
\end{align}
for three dielectric constants ($\epsilon_r = 4, 20, 80$) and Debye screening lengths ($\lambda$ = 25, 5, and 3 \AA), with corresponding bulk concentrations given in Table~I.

The spherical cavity has an additional parameter, the interpolation length $l$, that defines a range of smooth interpolation between the vacuum and solvent permittivity by means of a hyperbolic tangent function\cite{coons2016} and is hence similar to the $\Delta$ parameter of the error function in Eq.~(\ref{errfunc}) that we will use for all molecular solutes.
To obtain converged solvation free energies, we solved the linear PB equation for cubic boxes of eight different sizes ranging from 15 \AA \, to 50 \AA , adjusting the grid points to a uniform grid spacing of $l/3$.
We chose the grid spacing after a set of preliminary test calculations where this spacing proved to provide a proper interpolation in the transition region between vacuum and solvent permittivity.
The solute ion was placed in the center of these boxes. 
Charge neutrality --- cancellation of the integrated charge of the electrolyte ions and the solute ion --- is not guaranteed for finite box sizes but serves as a criterion for extrapolation.
We extrapolate the solvation free energy to that of a hypothetical box where charge neutrality is achieved.
The explicit extrapolation scheme differs slightly for each Debye length and is detailed in the Supporting Information.

The results are summarized in Table~I.
All in all, the mean absolute error (MAE) decreases with decreasing interpolation length $l$ as expected, i.e. when the numerical model approaches the analytical model where no interpolation region is necessary.
A charged molecule in a solvent of low dielectric permittivity combined with a low concentration of mobile ions is certainly the most challenging case for our implementation and extrapolation procedure.
Consequently, relative errors of almost 25~\% are observed in these cases.
We note here, that a spherical, charged solute is the worst case for our Cartesian grid with necessarily finite simulation box size and we expect the results obtained in this section to be an upper limit to the inaccuracies we expect for our implementation.

Another model system that is more comparable to the uncharged solutes that we discuss in the remainder of this article is a point dipole in spherical cavity that can analytically be described by the Kirkwood--Onsager model.
In the original article by Onsager\cite{onsager1936} only a point dipole in a spherical cavity surrounded by a dielectric was considered but Kirkwood\cite{kirkwood1939} extended this model to multipoles in electrolyte solution and provided analytical expressions for the solvation free energy.
The latter can be written as a sum of an electrolyte independent term $\Delta G_0$ and an electrolyte dependent term $\Delta G(\kappa)$:
\begin{align}
\Delta G = \Delta G_0 + \Delta G(\kappa) \, .
\end{align}

For a pure point dipole, these terms read
\begin{align}
\Delta G_0 = -\frac{\mu^2}{a^3}\frac{(\epsilon_r - 1)}{(2 \epsilon_r+1)}\, ,
\end{align}

which is the Onsager result with the electric dipole moment $\mu$, and 
\begin{align}
\Delta G(\kappa) =&- \frac{3}{2} \frac{\mu^2}{\epsilon_r a}\left[ \frac{\epsilon_r}{2\epsilon_r+1}\right]^2 \nonumber \\
&\times \frac{\kappa^2}{1+ \kappa a + \frac{1}{3}\kappa^2 a^2 + \frac{\epsilon_r-1}{2\epsilon_r+1}\frac{\kappa^2 a^2}{3}} \, .
\end{align}

We model the point dipole by two charges of $\pm 8e$ separated by $d=0.1$~\AA, such that $a>>d$ and $\mu=1.512$~D.
Since the solute does not contain charges, charge neutrality is fulfilled even for small box sizes and we can calculate the free energy of solvation directly from the largest box size considered (30~\AA) without extrapolation.
If no convergence was achieved with the standard damping parameters introduced above, results where taken from calculations with smaller box size (20~\AA \, or 15~\AA).

Table~II summarizes results for all combinations of the parameters $\epsilon = \{20,80\}$, $c^b=\{0.5,1.0\}$ mol/l, and $a=\{1,2\}$ \AA.
Again, the calculations converge to the analytic solution with decreasing interpolation length which brings the numerical model closer to the analytical one where such a region is absent.
Certainly, this comes at the price of a denser grid which can be optimized in future work by a more variable grid that is very fine in this interpolation region and more coarse everywhere else.
The need for a dense grid is more pronounced for solvents with a large dielectric permittivity because the electrolyte is more concentrated close to the boundary region between solute and solvent.
From the results for the Kirkwood--Onsager model, we conclude that  in our implementation the numerical accuracy can be as accurate as 0.01~kcal/mol without the need for extrapolation.

\begin{table}[tb]
  \begin{center}
    \caption{Electrostatic free energies for the Kirkwood--Onsager point dipole model. The values in the last three columns were calculated from the linear Poisson--Boltzmann equation for three different interpolation lengths $l$.}
    \label{tab:bohr}
    \begin{tabular}{llllll} 
    \hline
      $\epsilon_r$ &  \multicolumn{1}{c}{$c^b$} &\multicolumn{1}{c}{$\Delta G^\mathrm{exact}(\kappa)$} & \multicolumn{3}{c}{$\Delta G^\mathrm{LPBE}(\kappa)$}\\
      &    \multicolumn{1}{c}{[mol l$^{-1}$]} & \multicolumn{1}{c}{[kcal mol$^{-1}$]} & \multicolumn{3}{c}{[kcal mol$^{-1}$]}\\
       \multicolumn{3}{c}{} & $l = 0.50$ \AA & $l = 0.35$ \AA &$l = 0.25$ \AA  \\
      \hline
\multicolumn{6}{c}{$a=1$\AA} \\
20 & 0.5 & -0.514 & -0.595& -0.508 & -0.497 \\
20 & 1.0 & -0.866 & -0.952&-0.887 &-0.851 \\
80 & 0.5 & -0.042 & -0.101&-0.049 &-0.046 \\
80 & 1.0 & -0.076 & -0.160&-0.099 &-0.090 \\
\multicolumn{6}{c}{$a=2$\AA} \\
20 & 0.5 &-0.173 & -0.187&-0.181 &-0.176 \\
20 & 1.0 &-0.257 & -0.263& -0.262&-0.256 \\
80 & 0.5 &-0.017 & -0.052&-0.018 &-0.020 \\
80 & 1.0 &-0.021 & -0.066& -0.032&-0.033 \\
    \hline
    \multicolumn{2}{l}{\textbf{MAE}} & &0.051 & 0.010 &0.008    \\
    \multicolumn{2}{l}{\textbf{MRE} [\%]} && 88.3 & 14.3  & 13.6  \\
    \hline
    \end{tabular}
  \end{center}
\end{table}

\subsection{Parameters of the solvent models}
\label{sect:sechenov}
The PB solvent models introduced above include a number of parameters that need to be carefully adjusted in order to obtain accurate results.
These are the atom-specific radii $d_\alpha$ that define the solute cavity, the width parameter of the error function $\Delta$, the finite ion radii $R$ and the Stern-layer thickness $a$.
Motivated by the result of the previous section, that the interpolation width is not a crucial parameter as long as it is not too small, we keep $\Delta = 0.265$ \AA \, as in previous work.\cite{fisicaro2016}
We use experimental data on hydrated ions to set the finite ion radii $R$.
Since we are only concerned with sodium chloride solutions in this study, we employ an ion size of $R=4.3$ \AA, which is the average of the hydrated sodium ion size ($R=4.7$ \AA)\cite{kielland1937} and that of the hydrated chloride ion ($R=3.9$) \AA.\cite{kielland1937}

As is the case for other implicit solvation models like the polarizable continuum model (PCM), the proper selection of atom-specific radii $d_\alpha$ to construct the solute cavity is crucial.
This cavity determines the fraction of space where the solvent, characterized solely by its dielectric permittivity (Eq.~\ref{eq:dielectric}), resides, and, after inclusion of the additional Stern-layer parameter $a$, also the location of the mobile ions (Eq.~\ref{eq:lambda}).
Two choices are commonly employed for these atom-specific radii $d_\alpha$: van-der-Waals radii scaled by a factor of 1.2 (sVDW)\cite{bonaccorsi1984,klamt1993,herbert2016} and the solvent-accessible-surface (SAS),\cite{lee1971,tomasi2005} which is constructed by adding a solvent-dependent probe radius to the van-der-Waals radius
\begin{align}
d_\alpha^\mathrm{SAS} = d_\alpha^\mathrm{vdW} + d_\alpha^\mathrm{probe} \, ,
\end{align}
where in the case of water $d_\alpha^\mathrm{probe} = 1.4$ \AA, being half the distance to the maximum of the  O-O radial distribution function of pure water.\cite{lange2019,brookes2015}
The cavity created by the solvent-accessible surface is therefore usually larger than that of the scaled van-der-Waals radii which affects the total free energy of solvation but also its dependence on the electrolyte concentration $\Delta \Delta G_\mathrm{ion}$.
\begin{figure}[t]
\includegraphics[width=0.5\textwidth]{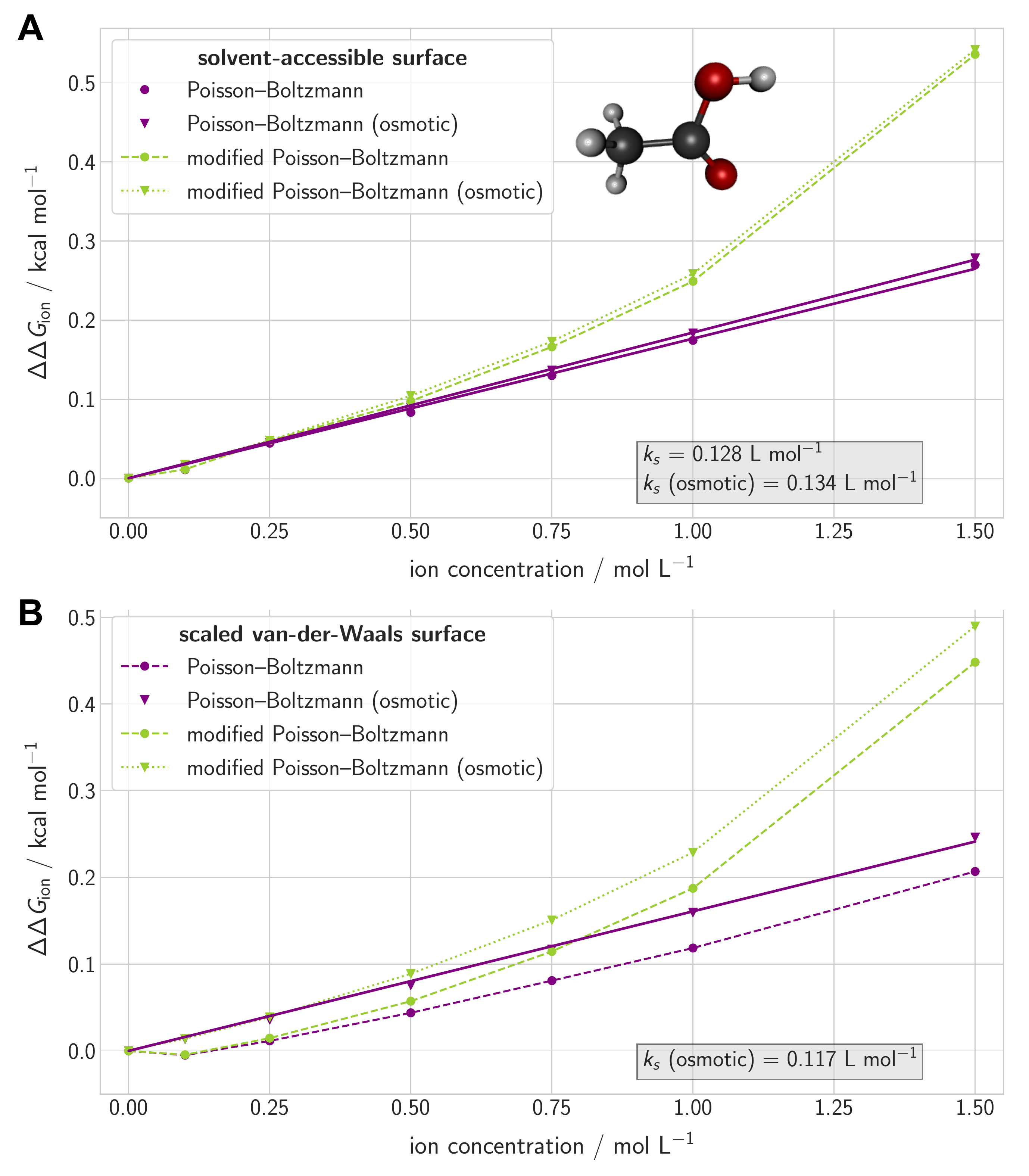}
\caption{Effect of the electrolyte concentration  on the free energy of solvation $\Delta \Delta G_\mathrm{ion}$ for acetic acid calculated for different electrolyte concentrations with the standard nonlinear PB model (purple) and the size-modified MPB model (green). The filled circles show the results obtained from the full expression, whereas the triangles indicate results obtained solely from the osmotic pressure term. Dashed and dotted lines are shown to guide the eye whereas the solid lines are the results of linear fits in order to obtain Sechenov coefficients. Panel~A shows the results for a solvent-accessible surface cavity, whereas results for the same set of calculations but with a scaled van-der-Waals cavity are displayed in panel~B.}
\label{fig:surfaces}
\end{figure}
The results in Figure~\ref{fig:surfaces} exemplify the differences observed for the two sets of atom-specific radii on the example of the acetic acid molecule.
As will be discussed in detail below, a linear dependence of the solvation free energy with the electrolyte concentration is observed experimentally, with the slope being described by the so-called Sechenov coefficient.
This linear behaviour is observed for the standard non-linear PB equation, but finite ion size annihilates the linearity.
Following previous authors,\cite{ringe2017} we concentrate on the standard PB equation in the following when determining a suitable parameter for the Stern-layer thickness, but will discuss this in more detail in a later section.
While for the SAS cavity hardly any difference is observed between the results obtained from the full PB expression compared to that from the osmotic pressure term only, this is not the case for the smaller sVDW cavity.
The slope is the same for larger concentrations but at low concentrations the purely electrostatic terms even cause a slight salting-in effect.
In the linear regime at larger concentrations, the slope depends only slightly on the parameterization of the cavity as can be seen from the similar Sechenov coefficients in panel A and B of Figure~\ref{fig:surfaces}.

In the following, we will employ the SAS in all further calculations as it provides the experimentally observed linear relation also for low electrolyte concentrations.
We have now determined all parameters of our model but one, with the Stern-layer thickness $a$ being optimized in the next section.

\subsection{Stern-layer thickness and Sechenov coefficients}

Since there is scarce experimental data on the exact thickness of the Stern-layer, especially around molecules rather than charged surfaces, we optimize this parameter in order to obtain the best possible agreement of our implicit solvent model with reliable experimental results.\cite{ringe2017}
The experimental data we can compare are the Sechenov coefficients $k_s$ introduced above,\cite{sechenov1892,long1952,ringe2017} that are directly calculated from the slope of the free energy of solvation with the electrolyte concentration.
The relation is simply
\begin{align}
k_s = \frac{\Delta \Delta G_\mathrm{ion}}{c^b} \frac{\mathrm{log}_{10}(e)}{k_\mathrm{B}T} \, .
\label{eq:sechenov}
\end{align}
Experimental data frequently shows a slight deviation from this linear data that is accounted for by an additional quadratic term.\cite{devisscher2018}
However, the corresponding second-order Sechenov coefficient is usually two orders of magnitude smaller and is therefore of minor importance for electrolyte concentrations in the range of up to 2~mol/L.
Additionally, the temperature dependence of the Sechenov coefficients is not accounted for in Eq.~\ref{eq:sechenov}, as the experimental data is obtained at around room temperature and the temperature dependence of the Sechenov coefficients is negligible between 5 and 30$^\circ$ C.\cite{may1978}
We discussed already in the previous section that this linear relation is only observed for the PB equation with point charges, whereas finite ion size leads to more pronounced salting-out effects for larger electrolyte concentrations.

In order to find an optimal Stern-layer thickness parameter $a$, we calculated the Sechenov coefficients for 39 molecules for which experimental data was available in Ref.~\citenum{endo2012}.
This set of molecule is our training set, whereas the 43 molecules from Ref.~\citenum{li2004} serve as our validation set.
Optimized molecular structures were taken from Ref.~\citenum{misin2016}.

While it is a general shortcoming of implicit solvent models that they cannot capture explicit solute-solvent interaction, this is especially severe when electrolyte solutions are considered.
Ion-pairing with electrolyte ions in zwitterionic molecules, structural changes upon cation-$\pi$ interaction, and strong bonding to very polar groups are either not at all or inadequately described.
Since we calculate the free energy of solvation for isolated molecules, any kind of solute-solute interaction is also neglected.
We therefore expect to observe a number of outliers in our training set (and, less problematic, validation set), that need to be identified and excluded to not bias the optimization procedure.
The random sample concensus (RANSAC) method,\cite{fischler1981} allows us to identify outliers and exclude them from the linear regression.
We used the \textsc{scikit-learn} implementation\cite{scikit-learn} of RANSAC in \textsc{python} with 1000 trials in order to ensure that all combinations of two data points required for the linear regression model were included, effectively eliminating the randomness of the sampling for our rather small dataset.
The residual threshold that defines the maximum residual for a data point to be counted as an inlier is another crucial parameter of the algorithm.
We decided to scan several values for the threshold parameter (0.1, 0.075, 0.05, 0.025, 0.01, 0.005, 0.001), where the loose threshold usually identified few or no outliers, whereas for the tightest thresholds almost all data points were identified as outliers.
\begin{figure}[t]
\includegraphics[width=0.5\textwidth]{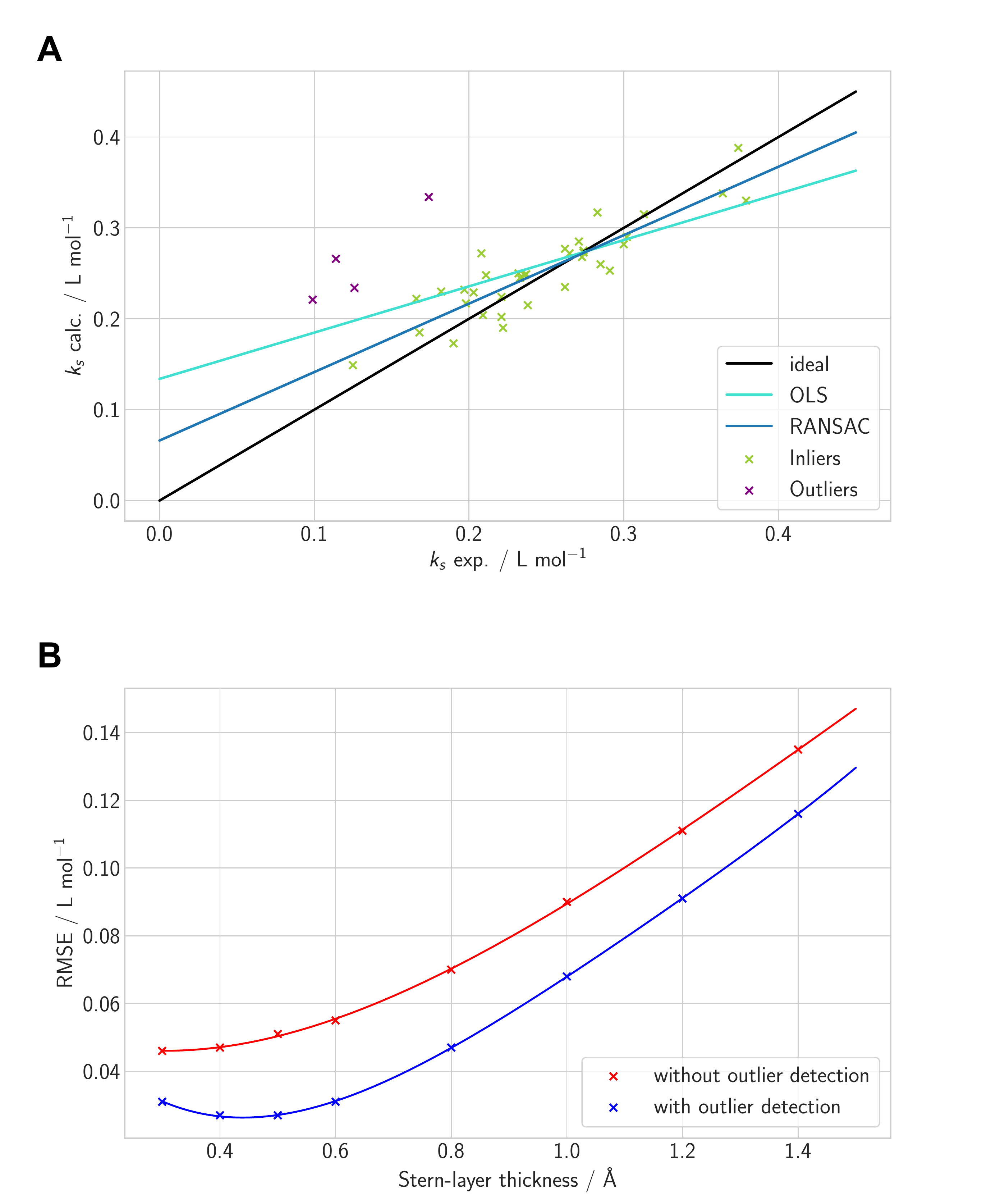}
\caption{Panel A: Example of the linear regression for the correlation between the experimental and calculated Sechenov coefficients $k_s$ for a Stern-layer thickness of $a=0.5$ \AA. The curves correspond to ideal correlation (black), ordinary least squares (OLS, light blue) and RANSAC with an error tolerance of 0.075 (dark blue), whereas the green crosses mark the inliers, and the purple crosses mark the outlying data points as identified by the RANSAC algorithm.\\
Panel B: Dependence of the RMSE between the calculated and experimental Sechenov coefficients with (blue) and without (red) exclusion of outliers. The minimum of the curve where outliers are excluded corresponds to a Stern-layer thickness of 0.44~\AA.}
\label{fig:ransac}
\end{figure}
Panel A of Figure~\ref{fig:ransac} shows an example of the linear regression for the Stern-layer thickness $a=0.5$ \AA.
The black line indicates perfect correlation between experimentally determined and calculated Sechenov coefficients, the light blue line is the result of an ordinary least squares fit (OLS), and the dark blue line results from the RANSAC algorithm with the threshold set to 0.075.
Clearly, the identification of four outliers significantly increases the coefficient of determination (RANSAC: $R^2=0.79$, OLS: $R^2=0.48$), and is also closer to ideal correlation.

After some initial trial calculations we decided on two criteria for the final RANSAC threshold selection (and, hence, the outlier detection):
\begin{enumerate}
\item The maximum number of outliers should not exceed one third of the full training set to allow for a meaningful fit.
\item A new set of outliers identified with a tighter threshold is only accepted if accompanied by a substantial increase in the coefficient of determination $R^2$. This is to avoid a large number of outliers are identified with only a marginal increase in the fit quality. We required $\Delta R^2 = 0.02$ per newly identified outlier. 
\end{enumerate}
These criteria were applied for calculated sets of Sechenov coefficients for eight Stern-layer thicknesses ($a = 1.4, 1.2, 1.0, 0.8, 0.6, 0.5, 0.4, 0.3$ \AA).
In all cases, the same four outliers were detected: bisphenol A, caffeine, 4-nitroanisole, 4-nitroaniline.
Bisphenol A was identified as an outlier also in a previous study\cite{misin2016}, where the deviation was attributed to an explicit interaction of the $\pi$-system with a sodium ions that induces structural changes.
In the case of caffeine, a (partial) dimerization might cause the large deviation between the experimental and calculated Sechenov coefficient.\cite{horman1984}
Why the aromatic nitro-compounds are outliers is difficult to speculate, especially since the agreement is much better for the two nitroalkanes 1-nitropentane and 1-nitrohexane.

We identified the optimal Stern-layer thickness parameter $a_\mathrm{opt}$, corresponding to the minimal root mean squared error (RMSE), by fitting a polynomial of fourth order to all data points obtained from the calculations on the training set with varying $a$ (see panel B of Figure~\ref{fig:ransac}).
For the curve (red) without outlier detection --- the OLS results --- no minimum can be observed, whereas for the curve with outlier detection (blue) --- the RANSAC results --- a minimum at $a_\mathrm{opt}= 0.44$ \AA \, can be identified, corresponding to an RMSE of 0.027 L mol$^{-1}$.
We hence recommend the thus optimized Stern-layer thickness in conjunction with the SAS for sodium chloride electrolyte solutions but note that the experimental data on the Sechenov coefficients is only of $m$L mol$^{-1}$ accuracy such that an RMSE that varies less than this contains no significant information.
The range of $0.4 < a < 0.5 $ \AA \, is therefore of indistinguishable accuracy and an equally suitable choice.

\begin{figure}[t]
\includegraphics[width=0.5\textwidth]{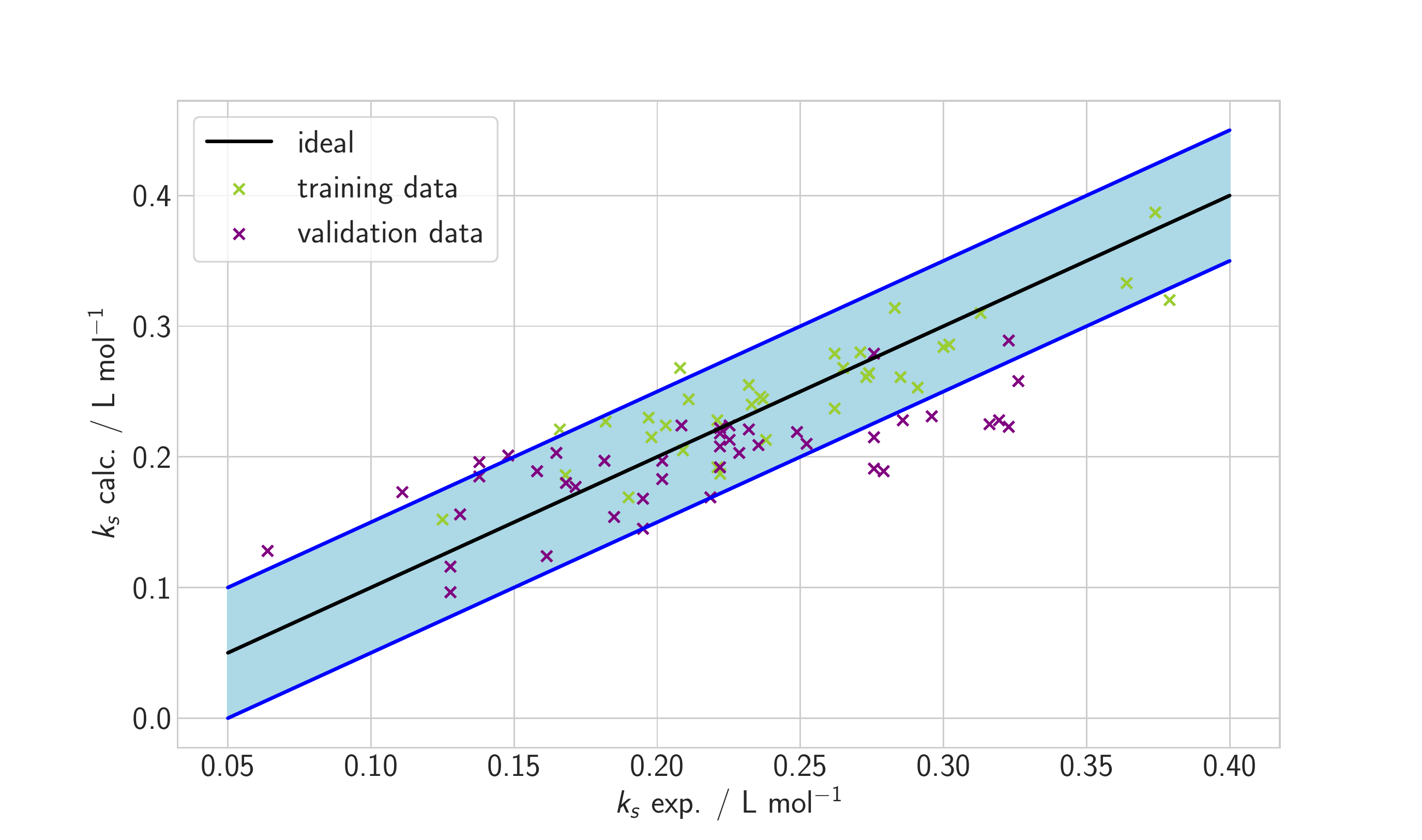}
\caption{Correlation between the experimental Sechenov coefficients and the calculated values obtained with the optimized Stern-layer thickness. Green markers indicate test data and purple markers indicate validation data. The black line corresponds to ideal correlation, whereas the blue area marks a spread of 0.1 L/mol.}
\label{fig:opt}
\end{figure}

In Figure~\ref{fig:opt} we show results obtained with the PB model employing the optimized Stern-layer thickness for the training (green markers) and validation set (purple markers).
The black line again corresponds to perfect correlation between theory and experiment and the blue area indicates a deviation band of 0.1 L/mol.
The RMSE for the validation set alone is 0.047 L mol$^{-1}$ and for the combined training (without outliers) and validation data it amounts to 0.039 L mol$^{-1}$.

In order to assess the quality of this result with other models, we rely on previous studies that certainly include different datasets but have substantial overlap with our data in all cases.
The COSMO-RS model\cite{klamt1995,klamt2000} shows a very high RMSE of 0.315 L mol$^{-1}$\cite{misin2016}, which is, however, due to a systematic overestimation because it can be reduced to 0.050 L mol$^{-1}$ when the empirical correction
\begin{align}
k_s = 0.335 k_s (\mathrm{COSMO}) + 0.060
\end{align}
is applied.\cite{endo2012}
The most meaningful comparison, however, is certainly the work of Ringe \textit{et al.}\cite{ringe2017}, since they use a similar PB model with a slightly different parameterization.
Their calculations with sodium chloride as electrolyte result in an RMSE of 0.068 L mol$^{-1}$ but it has to be noted that they do not employ any kind of outlier detection.
Excluding outliers will certainly reduce the RMSE significantly and bring their accuracy in the realm of our results.
To put these results into perspective, we compare to the polyparameter linear free energy relationships (pp-LFER) model.\cite{abraham1993,abraham2004} These highly parameterized calculations result in an RMSE of 0.047 L mol$^{-1}$ and are therefore comparable in accuracy to our model and that of Ringe \textit{et al.} but of no use for quantum-chemical calculations because the Sechenov coefficients are calculated directly without the electrostatic potential.

\subsection{Concentration dependence of the Stern-layer thickness}
For the optimization of the Stern-layer thickness parameter in the previous section we compromised on the PB model with point-like ions instead of considering the ion size or the influence of the hydration shell.
The linear dependence of the solvation free energy with the electrolyte dependence was only retained in this approximation and allowed us to fit to experimental data.
This, however, is a disappointing result from a conceptual perspective: a physically more involved model is abandoned in favor of a much more coarse approximation because experimental relations are not reproduced by the more elaborate model.
This non-linear behaviour can not be explained by a quadratic term in Eq.~\ref{eq:sechenov} because experimentally determined quadratic Sechenov coefficients are usually orders of magnitude smaller than the linear one and might even have a negative sign.\cite{devisscher2018}
In this section we argue that this lack of agreement is likely due to a neglect of another effect: the dependence of the Stern-layer thickness on the electrolyte concentration.

In a recent study, the dependence of the Stern-layer thickness on the electrolyte concentration was probed by X-ray photo-electron spectroscopy on colloidal silica nanoparticles in sodium chloride solution.\cite{brown2016}
The authors observed a steep decrease of the Stern-layer thickness at low concentrations, followed by a domain where it decreases linearly.
Taking into account both an electrolyte-concentration dependent Stern-layer thickness and finite size for the electrolyte ions in the MPB model can then yet again result in the linear Sechenov relation with the added benefit of less drastic approximations.

\begin{figure}[t]
\includegraphics[width=0.5\textwidth]{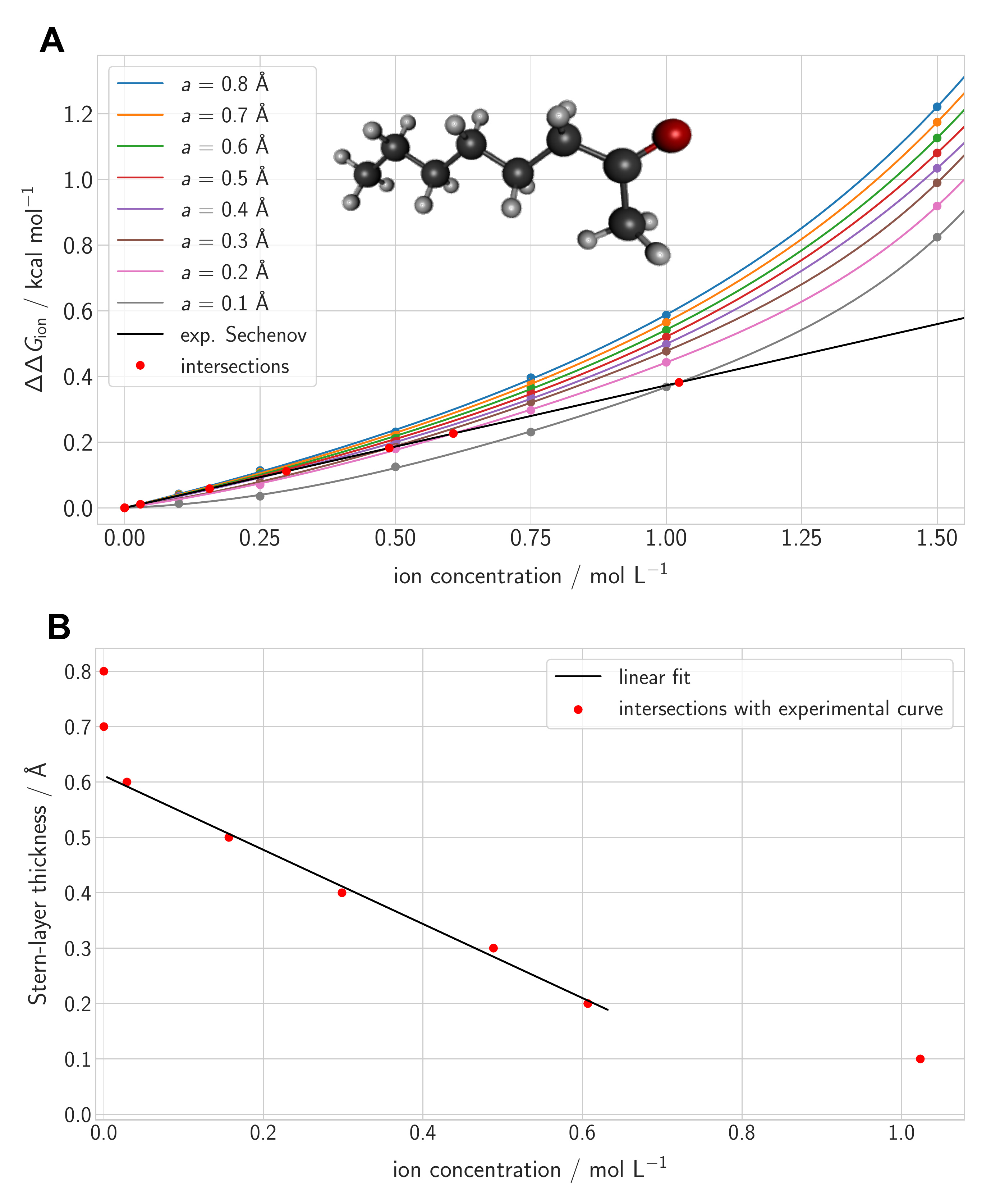}
\caption{Panel A: Electrolyte concentration dependence of the solvation free energy $\Delta \Delta G_\mathrm{ion}$ for 2-octanone calculated with the MPB equation for eight Stern-layer thicknesses $a$ ranging from 0.1~\AA \, to 0.8~\AA \, and corresponding experimental curve (black). The lines are polynomial fits to sixth order (see text). The red dots mark the intersections between the calculated and the experimental curves.\\
Panel B: Linear fit for the five intersection points between the experimental and calculated curves for the intermediate regime. In the absence of mobile ions (left), there is no need for a Stern-layer correction, whereas the linear dependence is lost for high ionic concentrations (right) in line with experimental results.}
\label{fig:stern}
\end{figure}
In panel A of Figure~\ref{fig:stern} we show the results of MPB calculations on 2-octanone for seven electrolyte concentrations and eight Stern-layer thicknesses between 0.1 \AA \, and 0.8~\AA.
The data points are fitted to a polynomial for proper interpolation (details on the polynomial fit are described in the SI).
The corresponding experimental curve (black line) obtained from the measured Sechenov coefficient is also shown as are the points (red) where the calculated lines intersect with the experimental one.
Assuming an otherwise adequate model --- most importantly an accurate parameter for the finite ion size --- these points mark the correct Stern-layer thickness for a given electrolyte concentration.
We then plot the Stern-layer thicknesses of these points as a function of the electrolyte concentration (see panel B of Figure~\ref{fig:stern}) and obtain a very similar pattern as the experimentally observed one.
At very low concentrations the concept of a Stern-layer is invalid but in the physiological saline ($\sim 0.15$ mol L$^{-1}$) and seawater regime ($\sim 0.6$ mol L$^{-1}$) a linear dependence is observed.
For even higher concentrations, the linear curve flattens out, which is to be expected because negative values for the Stern-layer thickness would be unphysical.
In conclusion, our results obtained from the comparison of the MPB calculations with the experimental curve are in line with the experimentally observed electrolyte concentration dependence of the Stern-layer thickness.

\begin{table}[h]
  \begin{center}
    \caption{Dependence of Stern-layer thickness $a$ on electrolyte ion concentration $c^b$ calculated from the intersections of the MPB curves with the experimental Sechenov curve for five molecules.}
    \label{tab:bohr}
    \begin{tabular}{llll} 
    \hline
    molecule & $k_s$ exp.\cite{endo2012}&$m = a/c^b$  & $a_\mathrm{max}$\\
     & [L mol$^{-1}$]&[\AA \, L mol$^{-1}$]& [\AA] \\
      \hline
      acetanilide & 0.197&-0.59 & 0.41 \\
      2-octanone & 0.273 & -0.67 &0.61\\
      111333-hexafluoro- & 0.222 &-0.90 & 1.02 \\
      2-propanol &&  \\
      $n$-propylbenzene & 0.262& -1.24 & 1.12 \\ 
      di-$n$-propylphtalate & 0.374& -1.42 & 0.80 \\
    \hline
    \end{tabular}
    \label{tab:stern}
  \end{center}
\end{table}

The slopes ($m$) and intercepts ($a_\mathrm{max}$) of these diagrams are compared for five molecules in Table~III.
While the slopes roughly correlate with the intercepts as might be expected, a correlation with the Sechenov constant or the size of the molecules is not evident.

The analysis of this section offers an explanation for the non-linear behaviour of the MPB model.
In practical calculations however, the application of the MPB model is hampered by the additional parameters that are hard to be determined from first principles.
These parameters are the electrolyte concentration dependence of the Stern-layer thickness and the finite ion size (pure ions or inclusion of some hydration).
If one of these parameters can be deduced from either experimental data (as for the Stern-layer thickness of the silica nanoparticles) or from molecular dynamics (MD) simulations, it will be possible to refit the remaining parameter with the aid of the experimental Sechenov coefficients to arrive at a more accurate implicit solvation model.
For now, however, in the absence of either experimental or MD data, we recommend the standard PB model with the optimized parameters described above.

\section{Conclusions}
In this article, we introduced our multigrid implementation of an implicit solvation model that includes the effect of electrolytes.
We describe the underlying Poisson--Boltzmann model with a hierarchy of approximations, namely linear versus non-linear Poisson--Boltzmann equations, inclusion of a Stern-layer correction and finite ion-size corrections.
Our derivation of the free energy of solvation including the effect of the electrolyte is based on the notion that these equations can be recast into an Euler-Lagrange equation.\cite{sharp1990}
We then challenged our implementation of the linear PB model by comparing to analytical solutions for the Born ion model and observed that the relative error for polar solvents and moderate to high salt concentrations is below 2~\%.
We then discussed the parameters of the model and optimized the Stern-layer thickness $a$ based on a fit to an experimentally observed linear relationship between the free energy of solvation and the electrolyte ion concentration which is described by so-called Sechenov coefficients.
For a cavity constructed as a solvent-accessible surface with a probe radius of 1.4~\AA \, for water, we arrived at $a_\mathrm{opt}=0.44$ \AA, by fitting to experimental data for 39 molecules and identifying outliers with the RANSAC algorithm.
The RMSE of 0.039~L/mol that we obtained by including results for a validation set of 43 molecules is comparable to that of previous work by Ringe~\textit{et al.} and as good as that obtained with the highly parameterized pp-LFER model.
In the final section, we attributed the failure of the size-modified PB model to a neglect of the electrolyte concentration dependence of the Stern-layer thickness.
We then demonstrated that in a certain concentration range, knowledge about this linear dependence can restore the experimentally observed linearity of the solvation free energy with electrolyte concentration.
We argued that this knowledge might be obtained either from experiment or from MD simulations, where the latter can also be informative on the finite ion size, i.e. the size of the hydration shell around the electrolyte ions.\\
To our knowledge, this is the first implementation of an implicit solvation model for electrolyte solutions in a Gaussian orbital based multi-purpose quantum chemistry program.
A related implementation\cite{ringe2016} based on numeric atom-centered orbitals can be found in the \textsc{FHI-aims}\cite{blum2009} package and for plane-wave basis sets for example in \textsc{VASPsol} (only linearized PB).\cite{mathew2014}\\
This enables us to investigate electrochemical processes for molecules or small cluster models with the inclusion of the effect of electrolyte solutions.
Future work will therefore focus on the implementation of nuclear gradients for the grid-based PB model so that solvation induced structural changes can be analyzed in addition to the change in the solvation free energy.
This will then also allow us to systematically study the effect of the linear vs. non-linear PB model and the influence of the finite ion size on electrocatalytic reactions.

\section*{Supplementary Material}
See supplementary material for a table demonstrating the convergence with grid density and examples of the extrapolation schemes for the Debye--H\"uckel model calculations.

\section*{Acknowledgements}
This material is based upon work supported by the U.S. Department of Energy, Office of Science, Office of Advanced Scientific Computing, and Office of Basic Energy Sciences, via the Scientific Discovery through Advanced Computing (SciDAC) program.\\
C. J. S. gratefully acknowledges financial support in form of an Early Postdoc.Mobility fellowship from the Swiss National Science Foundation.\\
Work by J.M.H. on the Poisson solver was supported by the National Science Foundation (CHE-1665322).\\
The authors declare the following competing financial interests: J. M. H and M. H.-G. are part owners of Q-Chem, Inc.
\section*{References}

\bibliographystyle{achemso}
\bibliography{references}

\end{document}